# Semiparametric Volatility Model with Varying Frequencies

*Jetrei Benedick R. Benito*
School of Statistics, University of the Philippines Diliman

*Joseph Ryan G. Lansangan*
School of Statistics, University of the Philippines Diliman

*Erniel B. Barrios*
ORCID:0000-0003-4209-2858
School of Statistics, University of the Philippines Diliman
corresponding author: ebbarrios@up.edu.ph

**Abstract**

In extracting time series data from various sources, it is inevitable to compile variables measured at varying frequencies as this is often dependent on the source. Modeling from these data can be facilitated by aggregating high frequency data to match the relatively lower frequencies of the rest of the variables. This however, can easily loss vital information that characterizes the system ought to be modelled. Two semiparametric volatility models are postulated to account for covariates of varying frequencies without aggregation of the data to lower frequencies. First is an extension of the autoregressive integrated moving average with explanatory variable (ARMAX) model, it integrates high frequency data into the mean equation (VF-ARMA). Second is an extension of the Glosten, Jagannathan and Rankle (GJR) model that incorporates the high frequency data into the variance equation (VF-GARCH). In both models, high frequency data was introduced as a nonparametric function in the model. Both models are estimated using a hybrid estimation procedure that benefits from the additive nature of the models. Simulation studies illustrate the advantages of postulated models in terms of predictive ability compared to generalized autoregressive conditionally heteroscedastic (GARCH) and GJR models that simply aggregates high frequency covariates to the same frequency as the output variable. Furthermore, VF-ARMA is superior to VF-GARCH since it exhibits some degree of robustness in a wide range of scenarios.

## 1. Introduction

Asset Price is a continuous-time continuous stochastic process (Tsay, 2002). The evolution of price can be described as a random walk but since it has no fixed level, it exhibits nonstationarity. The log difference of price (returns) is stationary series and satisfies the random walk hypothesis (Quigley, 2002). Random walk is based on the theory of efficient markets proposed by Fama (1970), stating that current price fully reflects all available information and is the only determinant of future price. Sufficient conditions for a market to be efficient is the absence of transaction costs, all information is readily available to all participants, and that all participants are in agreement on the implications of current information for current price and the distribution of future prices. While these requirements are obviously unrealistic it does not necessarily mean that the market is inefficient.

Market efficiency can be divided into three categories: weak, semi-strong, and strong market efficiency. The three categories are distinguished by how strict the conditions of market efficiency are satisfied. Weak efficiency assumes that historical data is enough to compute for the expected value of future prices. Semi-strong efficiency adjusts this assumption to include readily available public information such as trading volume or company disclosures. Strong efficiency assumes that all information (public and insider) is considered and reflected in the current prices. One possible source of market inefficiency can be the disagreement among investors on the implications of given information, this can lead to an increase in market volatility.

Volatility is the spread of the returns of a financial asset. It plays important roles in the estimation of the value of market risk, option pricing, risk management and portfolio management. Modeling volatility can improve the efficiency of parameter estimates and the accuracy of interval forecasts (Tsay, 2002). Despite its importance, forecasting volatility can be difficult since it is not directly observable. Each time point has only one observation, usually the closing price of the day, but the intraday volatility and variation between trading days are unobserved. This affects forecasting performance of conditional heteroscedastic models. Further challenge in forecasting volatility is proper accounting of the complex interconnections of financial indicators (Ladokhin, 2009).

We formulate a volatility model that incorporates financial data of varying frequencies. Specifically, a semiparametric model that provides robust estimates of target price levels and intervals for stop loss levels. The proposed model is intended to reduce the bias that traders can accrue in their analysis and assist in making objective, data-driven decisions. Model performance is evaluated against some statistical models like the Generalized Autoregressive Conditional Heteroscedastic (GARCH) and GJR-GARCH (Glosten et al, 1993) using both simulated and real-life financial data.

## 2. Modeling with Time Series Data

We summarize in this section some approaches in modeling time series data accounting for various stylized facts about the data.

### 2.1. Volatility

Conditional heteroscedasticity occurs when the variance of the output variable $\{y_t\}$ also depends on the variance of an input variable $\{x_t\}$, the implication being that changes to the variance of $\{x_t\}$, regardless of magnitude, will be magnified in the variance of $\{y_t\}$. Conditional heteroscedastic models can be one of two classes: a stochastic volatility model when it uses a stochastic equation to describe the dynamic variance $\sigma_t^2$, or a Generalized Autoregressive Conditional Heteroscedastic (GARCH) family of models when it uses an exact function to characterize the evolution of $\sigma_t^2$.

While volatility of an asset is not directly observable, it exhibits certain characteristics like clustering of time periods with presence of high and low volatility, it evolves continuously over time, it varies within a fixed range and is often stationary, and it reacts differently to large increases or decreases in price. The GARCH family of models is very popular as parsimonious descriptions of a range of empirical stylized facts (Blasques et al., 2016). Initially introduced from the works of Engle (1982) and Bolleslev (1986), GARCH models elegantly capture volatility clustering. GARCH models are able to account for volatility clustering in financial data and also provides a parametric function that describes the evolution of volatility. GARCH however, assumes that positive and negative shocks have the same effects on volatility. But this is not the case since price reacts differently to positive and negative shocks because of how traders interpret them variably.

The shocks or innovations, $a_{t-1}$, in a GARCH model are referred to as "news". Yu (2012) noted that volatility of equities responds asymmetrically to news, bad news has a different impact on future volatility compared to good news of similar magnitude. The asymmetry is most apparent during a market crisis when volatility is high. Wu and Xiao (2002) proposed that the asymmetry could be due to changes in financial leverage of securities as a response to changes in the value of equity. Due to this asymmetry, Glosten et al. (1993) proposed an extension of the GARCH model that incorporates a dummy variable in the variance equation to differentiate periods of negative and positive shocks. The model is of the form:

$$\sigma_t^2 = \omega + (\alpha + \gamma I_{t-1})a_{t-1}^2 + \beta \sigma_{t-1}^2 \tag{1}$$

where $I_{t-1} = \begin{cases} 1, \text{ when } a_{t-1} < 0 \\ 0, \text{ otherwise} \end{cases}$, $\gamma$ is the leverage effect.

Equation (1) referred to as the GJR-GARCH model in this paper is a threshold GARCH that assumes the impact of a random shock $a_{t-1}^2$ on the conditional variance $\sigma_t^2$ is different when $a_{t-1}$ is positive than when $a_{t-1}$ is negative. Glosten et al. (1993) noted that while conditional variance persists in high frequency data, persistence is minimal in low frequency (i.e., monthly-level) data.

Engle and Ng (1993) introduced the news impact curve as a measure of the effect of the asymmetry, confirming the findings of Glosten et al. (1993) and concluded that the GJR-GARCH was the best parametric model. Engle and Ng (1993), in turn, proposed a partially non-parametric $ARCH(\infty)$ model (PNP).

Typically, a time series model involves data obtained at the same frequency. However, it is possible to encounter a situation where the data will be of mixed frequency, commonly occur when data are extracted from different sources.

## 2.2. Data with Mixed Frequency

Modeling data of mixed frequencies is growing interest since more observable data is becoming available at different and higher frequencies. Macroeconomic indicators are an example of mixed frequency data, with inflation rate reported at a monthly level while it is quarterly for GDP. Low frequency variable may be driven by the movement of relevant information measured at higher frequencies. The analysis of mixed frequency data enhances the potential for further gains in forecasting performance. However, it will introduce complexities in the estimation of the model. Ghysels (2016) noted that analysis at common low frequencies and sampling from high frequency data can easily lead to misspecification errors of the co-movements among the variables. It was further noted that various parsimonious parametrizations inspired by mixed data sampling (MIDAS) regression characterizes the misspecification of a traditional common low frequency VAR. Ghysels et al (2007) observed that while MIDAS focused on volatility modeling with mixed frequency, it has far more applications in various areas like macroeconomics and finance. Some new extensions of the MIDAS framework were developed and empirical work on microstructure noise and volatility forecasting presented. However, Kuzin et al (2011) attempted to compare parsimonious

MIDAS and mixed frequency VAR (without restriction on the dynamics, and hence, easily attract the curse of dimensionality) but realized difficulties in a priori comparison and concluded that this can only be done empirically.

Mariano and Ozmucur (2015) proposed a high-frequency forecasting model for GDP growth that used macroeconomic indicators of varying frequencies. The model takes the form of dynamic time series models that combined latent factors with the mixed frequency data. Malabanan, et al (2021) proposed a semiparametric spatiotemporal model that incorporated covariates of varying frequencies. No specific functional form is assumed between high frequency data and the response variable, influence is accounted in terms of a nonparametric function. The model have better predictive ability than ordinary generalized additive models that aggregates high frequency data to match with frequency of the response variable (relatively lower frequency).

## 2.3. Generalized Additive Model

Hastie and Tibshirani (1986) introduced the generalized additive model (GAM), an extension of the generalized linear model that replaces the linear predictor $\eta = \sum_j \beta_j x_j$ with the smoother additive predictor $\eta = \sum_j s_j(x_j)$. The regression model is then expressed in a more general model will form in Equation (2).

$$E(Y|X_1, \ldots, X_p) = s_0 + \sum_{j=1}^{p} s_j(x_j) \qquad (2)$$

where the smooth functions, $s_j(\cdot)$, are standardized so that $E[s_j(x_j)] = 0$. GAM is more flexible than its GLM counterpart since the former are estimated using scatterplot smoothers, which are nonparametric, while the latter are estimated through likelihood-based parametric methods. GLM is contrained with accuracy of the postulated linear model to capture the dynamics reflected in the data generating process.

Assuming that the model

$$Y = s_0 + \sum_{j=1}^{p} s_j(x_j) + \varepsilon \qquad (3)$$

is correct, and $s_0, s_1(\cdot), \ldots, s_{j-1}(\cdot), s_{j+1}(\cdot), \ldots, s_p(\cdot)$ are known, define the partial residuals as

$$R_j = Y - s_0 - \sum_{k=j} s_k(x_k) \qquad (4)$$

This follows that $E(R_j|X_j) = s_j(x_j)$ and minimizes

$$E\left(Y - s_0 - \sum_{k=1}^{p} s_k(x_k)\right)^2 \qquad (5)$$

Buja et al. (1989) proposed that while the $s_j(\cdot)$'s are unknown they can be estimated using linear smoothers through the iterative procedure known as the backfitting algorithm given below:

1) Initialization:

Let $s_0 = E(Y)$, $s_1^1(\cdot) \equiv s_2^1(\cdot) \equiv \cdots \equiv s_p^1(\cdot) \equiv 0$, $m = 0$

2) Iteration:

   $m = m + 1$

   For $j = 1$ to $p$ do:

   $$R_j = Y - s_0 - \sum_{k=1}^{j-1} s_k^m(X_k) - \sum_{k=j+1}^{p} s_k^{n-1}(X_k) \quad (6)$$

   $$s_j^m(X_j) = E(R_j|X_j) \quad (7)$$

3) Convergence: When
   $$RSS = E\left(Y - s_0 - \sum_{j=1}^{p} s_j^m(x_j)\right)^2 \quad (8)$$
   exhibit minimal changes in subsequent iteration.

Buja et al. (1989) highlighted the practicality of the algorithm in fitting GAM with linear smoothers. Opsomer (2000) examined the case of local polynomial regression smoothers and found that the estimators achieve the same rate of convergence as those of univariate local polynomial regression in the backfitting algorithm.

## 2.4. Semiparametric Modeling

Powell (1994) defines the semiparametric model to be a hybrid of parametric and nonparametric functions in postulating and fitting statistical models. This combines the parametric form of the components of the data generating process with the weak nonparametric restrictions. Often, the parametric components represent the relationship between the covariates and response variables while the nonparametric component represents the distribution of the unobservable errors. Andrews (1994) established consistency and asymptotic normality of a wide range of semiparametric estimators that optimizes iteratively an objective function that is initialized with infinite-dimensional nuisance estimator. Even when the true parameters lies on the boundary of the parameter space, Andrews (1999) proved that such asymptotic properties of the extremum still holds.

As a hybrid model, semiparametric models have the advantages and disadvantages of their parametric and nonparametric counterparts. On one hand, semiparametric estimators are consistent under a broader range of conditions compared to those from a parametric approach; and the former are more precise as opposed to nonparametric estimators. On the other hand, correctly specified maximum likelihood estimators are more efficient and semiparametric estimators are still sensitive to misspecification.

## 2.5. Semiparametric Volatility Modeling

Nonparametric GARCH models have two issues: they suffer from the curse of dimensionality, i.e., as the number of dimensions increases, the optimal rate of convergence decreases, and; they are difficult to interpret. Additive models offer a flexible parsimonious alternative. While parametric GARCH models avoid those issues raised against nonparametric GARCH, the model needs to be correctly specified. The classical assumption of conditional normality is violated in many empirical data, leading to misspecification errors further contributing to losses in statistical efficiency (Blasque et al., 2016). Semiparametric GARCH models were proposed to avoid the loss of efficiency due to misspecification error.

Wu and Xiao (2002) proposed a semiparametric model that used the innovations as a general function of the market volatility index and used kernel estimation for the unknown functions. Linton and Mammen (2005) expanded on the PNP model proposed by Engle and Ng (1993). The former employed an estimation method based on kernel smoothing and profiled likelihood. The PNP nests the $GARCH(1,1)$ but permits a more general functional form that allows for asymmetry. This model found evidence of asymmetric news impact in simulated and actual data on the S&P 500.

## 3. Semiparametric Model with Varying Frequencies

Consider two stationary time series $\{y_t\}$ and $\{v_t\}$ following an ARMAX process in Equation (9),

$$\phi_p(B)y_t = \theta_q(B)a_t + \psi(B)v_t \tag{9}$$

where $y_t$ is the response variable at time t, $v_t$ is the covariate at time t. Following the definition used in Campano (2012), the innovations $a_t$ behave as $a_t = \sigma_t \varepsilon_t$ where $\varepsilon_t \sim N(0,1)$ and $\sigma_t$ is a sequence of volatilities that are a function of past realizations. The innovations are also assumed to be asymmetric. The sequence of volatilities are postulated to follow the GJR-GARCH is Equation (10):

$$\sigma_t^2 = \omega + (\alpha + \gamma I_{t-1})a_{t-1}^2 + \beta \sigma_{t-1}^2 \tag{10}$$

where $I_{t-1} = 1$ when $a_{t-1}<0$. Consider for example, in financial trading, the variables $\{y_t\}$ and $\{v_t\}$ represent the weekly returns of an asset and its weekly percentage change in trading volume, respectively. The parameter $\phi$ is the effect of the previous weeks' returns to current returns, while $\psi$ is the effect of the weekly percentage change in trading volume.

### 3.1. Additive Models with Varying Frequency

Two models are proposed in this paper. The higher frequency data was not aggregated in order to preserve the information contained and incorporate its effect on the volatility and mean, respectively. The higher frequency covariate, $x_{i,t-1}$, represents for example the daily price movement of the previous week.

The first postulated model is an extension of Equation (10), to accommodate the higher frequency covariate by adding a nonparametric component $f_i(\cdot)$ into the variance equation given in Equation (11).

$$\sigma_t^2 = \omega + (\alpha + \gamma I_{t-1})a_{t-1}^2 + \beta\sigma_{t-1}^2 + \sum_{i=1}^{5} f_i(x_{i,t-1}) \quad (11)$$

where $x_{i,t-1}$ is a covariate measured at a higher frequency than the response variable.

The second postulated model is an extension of Equation (9). The nonparametric component is instead added into the mean equation given in Equation (12).

$$\phi_p(B)y_t = \theta_q(B)a_t + \psi(B)v_t + \sum_{i=1}^{5} f_i(x_{i,t-1}) \quad (12)$$

The proposed models combining Equation (9) and Equation (11), and extending Equations (9) and (10) with Equation (12) are referred to as the varying frequency general autoregressive conditionally heteroscedastic (VF-GARCH) and varying frequency autoregressive moving average (VF-ARMA), respectively.

### 3.2. Estimation Procedure

Both models are estimated through the modified backfitting algorithm. The iterative process in both procedures ends when they reach convergence in terms of MSE.

### 3.2.1. Estimation Algorithm for VF-GARCH

Given the model presented in Equations (9) and (11), the parameters of the mean $\{\hat{\phi}, \hat{\theta}, \hat{\psi}\}$ equation are estimated first, then the parametric components of the volatility model $\{\hat{\omega}, \hat{\alpha}, \hat{\beta}, \hat{\gamma}\}$, and finally, the nonparametric component $f(\cdot)$ that captures the contribution of the high frequency data. The hybrid estimation algorithm follows:

Step 1: Fit $\{y_t\}$ and $\{v_t\}$ into an ARMAX model to obtain the initial estimates for $\{\phi, \theta, \psi\}$ as well as the residuals and fitted values using the equation: $\hat{y}_{t,ARMAX} = y_{t,orig} - e_{t,ARMAX}$.

Step 2: Fit $e_{t,ARMAX}$ into a GJR-GARCH Model to obtain the initial estimates for $\{\omega, \alpha, \beta, \text{and } \gamma\}$ and the residuals, $e_{t,GJR}$, and the fitted values, $\hat{y}_{t,GJR}$

Step 3: Fit $e_{t,GJR}$ into a GAM, estimating the nonparametric components via spline smoothing. Compute for the residuals, $e_{t,GAM}$, and the fitted values, $\hat{y}_{t,GAM}$

Step 4: After Step 3, there is now a smoothing function to estimate the effect of $x_{it-1}$ as well as parameter estimates for $\hat{\phi}, \hat{\theta}, \hat{\psi}, \hat{\omega}, \hat{\alpha}, \hat{\beta}, \text{and } \hat{\gamma}$. Update the dependent variable using the equation: $y_{t,new} = y_{t,orig} - e_{t,GAM}$

Step 5: The algorithm iterates from Step 1 using the updated response computed in Step 4. In Step 4, the response variables are updated using the original values and the latest computed residuals from Step 3. The iteration continues until there is minimal changes in the MSE (e.g., MSE<0.005). The fitted values of the proposed model is the sum of the fitted values from the previous steps: $\hat{y}_{t,Model1} = \hat{y}_{t,ARMAX} + \hat{y}_{t,GJR} + \hat{y}_{t,GAM}$

This estimation algorithm follows the framework of multiple timeframe analysis. The parameters in the low frequency data were estimated first to provide an overview of the series, the nonparametric components were estimated afterwards in order to adjust the initial estimates. The procedure is repeated until it reaches a convergence in the MSE.

### 3.2.2. Estimation Algorithm for VF-ARMA

Given the model presented in Equations (12) and (10), the nonparametric component is first estimated. The response variables are then updated and used to estimate the parametric components of the mean $\{\hat{\phi}, \hat{\theta}, \hat{\psi}\}$, and finally, the parameters of the volatility model $\{\hat{\omega}, \hat{\alpha}, \hat{\beta}, \hat{\gamma}\}$. The algorithm is as follows:

Step 1: Fit $\{y_t\}$ and $\{x_{it-1}\}$ into a GAM, estimating the nonparametric components via spline smoothing. Compute for the fitted values, $\hat{y}_{t,GAM}$, and the corresponding residuals. Update the dependent variable using the equation: $y_{t,updated} = y_{t,orig} - e_{t,GAM}$

Step 2: Fit $\{y_{t,updated}\}$ and $\{v_t\}$ into an ARMAX to obtain the initial estimates for $\{\phi, \theta, \psi\}$ as well as the residuals, $e_{t,ARMAX}$. Compute for the fitted values using the equation: $\hat{y}_{t,ARMAX} = y_{t,orig} - e_{t,ARMAX}$

Step 3: Fit $e_{t,ARMAX}$ into a GJR-GARCH Model to obtain the initial estimates for $\{\omega, \alpha, \beta, \text{and } \gamma\}$ and the residuals, $e_{t,GJR}$, and the fitted values, $\hat{y}_{t,GJR}$

Step 4: After Step 3, there is now a smoothing function to estimate the effect of $x_{it-1}$ as well as parameter estimates for $\hat{\phi}, \hat{\theta}, \hat{\psi}, \hat{\omega}, \hat{\alpha}, \hat{\beta}, \text{and } \hat{\gamma}$. Compute for a new dependent variable using the original values and the residuals from step 3 using the equation: $y_{t,new} = y_{t,orig} - e_{t,GJR}$

Step 5: The algorithm iterates from Step 1 using the new values for the response computed in Step 4. In Steps 1 and 4, the response variables are updated using the original values. The iteration continues until there is minimal changes in the MSE (e.g., MSE<0.005). The fitted values of the proposed model will be the sum of the fitted values from the previous steps: $\hat{y}_{t,Model1} = \hat{y}_{t,GAM} + \hat{y}_{t,ARMAX} + \hat{y}_{t,GJR}$

This estimation algorithm follows the framework of semi-strong market efficiency. The nonparametric components were estimated first because they represent the past information of the response variable. The parametric components were then estimated as they represent any additional information related to the response. The algorithm proceeds until it reaches convergence in terms of MSE.

## 3.3. Simulation Studies

The proposed models are evaluated using simulated data. The high frequency data was first generated using a modified form of the model for continuously compounded returns in Equation (13).

$$x_i = x_0 e^{\sum_{j=1}^{i} s_j} - \psi u_i, i = 1, \dots, T \qquad (13)$$

where $s_j$ is a series generated from $N(\mu_{daily}, \sigma_{daily}), j = 1, \dots, T$

$u_i$ is a series generated from $U(-1,1)$

$\psi$ is the weight of $u_i$

$x_0$ is a preset value

The covariate $\{x_i\}$ and the series $u_i$ are then sequentially aggregated into groups of five. The covariate $\{v_t\}$ is then computed from $u_i$ from the following procedure:

$$w_{it} = 1000000 * |u_{it}|; i = 1,...,5 \qquad V_t = \sum_{i=1}^{5} w_{it} \qquad v_t = \ln \frac{V_t}{V_{t-1}}$$

The response variable is computed using Equation (14):

$$y_t = \ln \frac{x_{5,t}}{x_{5,t-1}} \tag{14}$$

Different values for drift and volatility for low frequency ($\mu_{annual}$ and $\sigma_{annual}$) data (e.g., annual) were considered. Volatility was assumed to be directly proportional to the drift, values ranged from the cases with low volatility levels up to the opposing extreme with high volatility levels. The values were converted from low frequency levels, $\mu_{annual}$ and $\sigma_{annual}$, into high frequency (e.g., daily) levels, $\mu_{daily}$ and $\sigma_{daily}$ from $\mu_{daily} = \frac{\mu_{annual}}{252}$ and $\sigma_{daily} = \frac{\sigma_{annual}}{\sqrt{252}}$. Different weights of the covariate $u_i$ were considered, from balanced ($\psi = 50\%$) to unbalanced ($\psi = 20\%$ or $70\%$).

The series was also simulated at different lengths. The proposed models were evaluated for a short series, where $T = 255$ or 51 observations in the lower frequency, which is equivalent to one trading year. A length of 510 observations, equivalent to two trading years, was also considered. Finally, the performance of the model was also evaluated in the long term, where $T = 1530$, equivalent to six trading years.

The presence and absence of dependencies in the high frequency data were also considered. The functional form of the high frequency data was also allowed to vary between linear and exponential. Table 1 summarizes the simulations settings considered in this paper.

**Table 1. Simulation Settings**

| Parameter | Parameter Settings | Number of Settings |
|---|---|---|
| Length of series | a. $T = 255$<br>b. $T = 510$<br>c. $T = 1530$ | 3 |
| Drift and Volatility | a. $\mu_{annual} = 20\%$ and $\sigma_{annual} = 30\%$<br>b. $\mu_{annual} = 40\%$ and $\sigma_{annual} = 45\%$<br>c. $\mu_{annual} = 80\%$ and $\sigma_{annual} = 60\%$ | 3 |
| Weight of $u_i$ | a. $\psi = 20\%$<br>b. $\psi = 50\%$<br>c. $\psi = 70\%$ | 3 |

| Dependencies in High Frequency Data | a. Independent<br>b. Autocorrelated ($\emptyset = 0.5$) | 2 |
|---|---|---|
| Functional form of High Frequency Data | a. Linear: $x_0 e^{\sum_{j=1}^{i} s_j} - \psi u_i$<br>b. Exponential: $x_0 e^{\sum_{j=1}^{i} s_j - \psi u_i}$ | 2 |

## 4. Results and Discussion

The postulated models along with the estimation algorithms were assessed using simulated and actual data. A total of 108 scenarios were considered. Discussion of the analysis is divided into sections concerning the high frequency data: (1) varying frequency in the mean, (2) varying frequency in the variance, (3) presence of dependencies, and (4) varying functional form. The performance of the models is assessed in terms of the root mean squared error (RMSE) and the mean absolute deviation (MAD). The standard parametric GARCH and GJR-GARCH models with aggregation on the high frequency variables served as benchmark models. The postulated models were also validated using out-of-sample forecasts based on the median absolute percentage error (MdAPE).

Figures (1) to (3) show the sample plots of the high frequency data for the first replicate of the simulated data. The series exhibited a trend when the high frequency data is linear. The range of the series is larger when the high frequency data is autocorrelated. As the length of the series increased, the range of an exponential series was greater than a linear series, regardless of whether there are dependencies or not in the high frequency data.

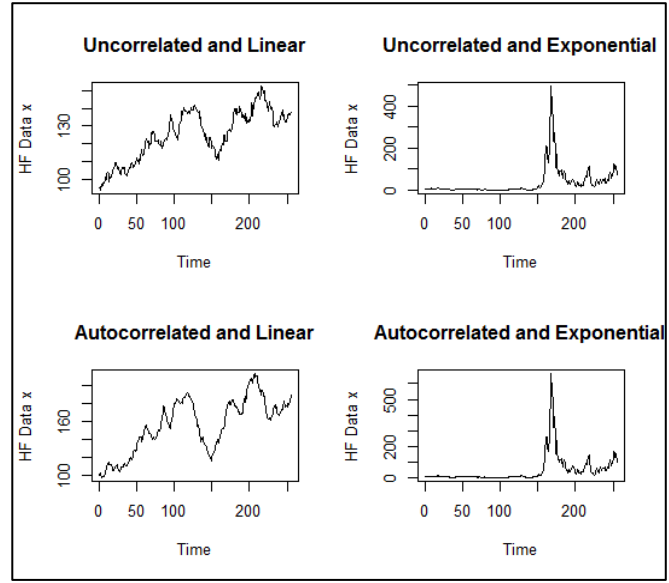

**Figure 1. Sample Graph of the High Frequency Data**
($T = 255$, $\mu_{annual} = 20\%$, $\sigma_{annual} = 30\%$, $\psi = 50\%$, *first replicate*)

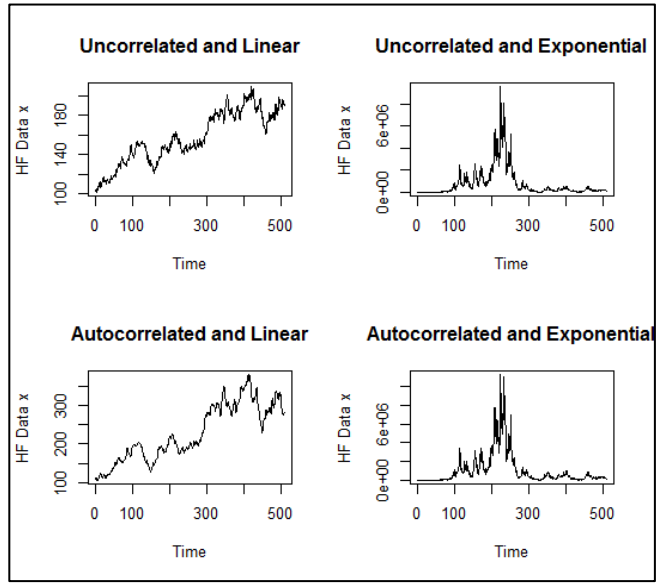

**Figure 2. Sample Graph of the High Frequency Data**
($T = 510$, $\mu_{annual} = 20\%$, $\sigma_{annual} = 30\%$, $\psi = 50\%$, *first replicate*)

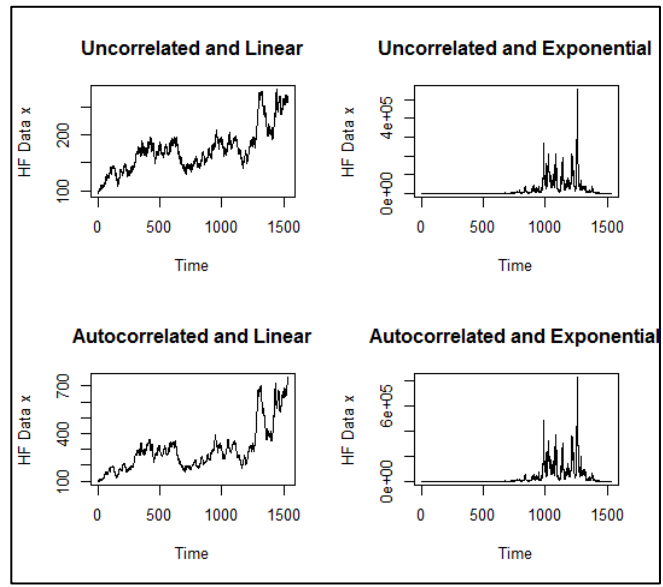

**Figure 3. Sample Graph of the High Frequency Data**
($T = 1530$, $\mu_{annual} = 20\%$, $\sigma_{annual} = 30\%$, $\psi = 50\%$ *first replicate*)

## 4.1. Varying Frequency in the Mean

Table 2 shows the average root-mean-square error and mean absolute deviation for all scenarios when $\mu_{annual} = 20\%$ and $\sigma_{annual} = 30\%$ and $\psi = 50\%$. The performance of the model improves as the length of the series increases. However, in the long series, i.e.: T = 1530, the proposed model failed to reach convergence for 50% of its replicates when the high frequency data is uncorrelated. This is an unlikely scenario since high frequency data usually exhibits volatility, counter intuitive to the uncorrelatedness assumption. Take note that when the high frequency data is exponential in form, the VF-ARMA still yields the lowest average RMSE and MAD.

**Table 2. RMSE and MAD for VF-ARMA and Benchmark Models**

|        |       | T    | VF-ARMA |        | GARCH  |        | GJR    |        |
|--------|-------|------|---------|--------|--------|--------|--------|--------|
|        |       |      | RMSE    | MAD    | RMSE   | MAD    | RMSE   | MAD    |
| Linear Form | (a)* | 255  | 0.2188  | 0.0180 | 0.4015 | 0.0323 | 0.4015 | 0.0323 |
|        |       | 510  | 0.1578  | 0.0131 | 0.4194 | 0.0336 | 0.4194 | 0.0336 |
|        |       | 1530 | 0.1023  | 0.0086 | 0.4250 | 0.0339 | 0.4250 | 0.0339 |
|        | (b)*  | 255  | 0.4056  | 0.0323 | 0.6833 | 0.0550 | 0.6833 | 0.0550 |
|        |       | 510  | 0.3071  | 0.0243 | 0.7097 | 0.0568 | 0.7097 | 0.0568 |
|        |       | 1530 | 0.2257  | 0.0183 | 0.7160 | 0.0571 | 0.7160 | 0.0571 |
| Exponential Form | (a)* | 255  | 3.5483  | 0.2778 | 4.5495 | 0.3660 | 4.5495 | 0.3660 |
|        |       | 510  | 2.5064  | 0.1813 | 5.1511 | 0.4100 | 5.1511 | 0.4100 |
|        |       | 1530 | 1.6751  | 0.1124 | 5.7437 | 0.4583 | 5.7437 | 0.4583 |
|        | (b)*  | 255  | 3.4763  | 0.2704 | 4.6142 | 0.3681 | 4.6142 | 0.3681 |
|        |       | 510  | 2.5272  | 0.1858 | 5.3220 | 0.4272 | 5.3220 | 0.4272 |
|        |       | 1530 | 1.6328  | 0.1119 | 5.7625 | 0.4604 | 5.7625 | 0.4604 |

*For cases where $\mu_{annual} = 20\%$, $\sigma_{annual} = 30\%$, and $\psi = 50\%$*
*(a) the high frequency data is uncorrelated; (b) the high frequency data is autocorrelated*

Table 3 shows the number of times the VF-ARMA failed to achieve converge. Note that as the length of the series increases, the number of occurrences of divergence also increases. In a shorter series (T = 255 and 510), this is negligible as it only constitutes less than 10% of the replicates per scenario. It becomes an issue in the long series, i.e.: T = 1530, where at most 47 of the 100 replicates failed to reach convergence. Although as drift increases, the occurrences will decrease from at most 47 (when $\mu_{annual} = 20\%$) to 22 (when $\mu_{annual} = 80\%$). This illustrates an issue with VF-ARMA, it encounters problems with convergence when the series is long and the high frequency data is uncorrelated and linear in form. Interestingly, this is not an issue when the high frequency variable is autocorrelated, a more common characteristic of high frequency time series.

**Table 3. Number of times the VF-ARMA Failed to Achieve Convergence**

|      |        | $\psi$ | T = 255 |             | T = 510 |             | T = 1530 |             |
|------|--------|--------|---------|-------------|---------|-------------|----------|-------------|
|      |        |        | Linear  | Exponential | Linear  | Exponential | Linear   | Exponential |
| (1)* | (a)**  | 20%    | 0       | 1           | 5       | 0           | 47       | 6           |
|      |        | 50%    | 1       | 1           | 8       | 4           | 47       | 9           |
|      |        | 70%    | 2       | 2           | 7       | 3           | 44       | 9           |
|      | (b)**  | 20%    | 0       | 1           | 1       | 2           | 1        | 3           |
|      |        | 50%    | 1       | 4           | 1       | 2           | 0        | 5           |
|      |        | 70%    | 1       | 0           | 0       | 6           | 0        | 12          |
| (2)* | (a)**  | 20%    | 1       | 0           | 3       | 5           | 22       | 2           |
|      |        | 50%    | 0       | 3           | 2       | 3           | 22       | 6           |
|      |        | 70%    | 0       | 1           | 2       | 5           | 19       | 7           |
|      | (b)**  | 20%    | 2       | 1           | 1       | 2           | 1        | 8           |
|      |        | 50%    | 1       | 1           | 0       | 5           | 1        | 7           |
|      |        | 70%    | 1       | 2           | 0       | 8           | 1        | 12          |
| (3)* | (a)**  | 20%    | 1       | 3           | 3       | 5           | 18       | 6           |
|      |        | 50%    | 2       | 0           | 3       | 4           | 21       | 10          |
|      |        | 70%    | 1       | 1           | 3       | 4           | 19       | 9           |
|      | (b)**  | 20%    | 3       | 4           | 0       | 2           | 2        | 4           |
|      |        | 50%    | 2       | 2           | 0       | 5           | 3        | 4           |
|      |        | 70%    | 5       | 1           | 2       | 7           | 2        | 10          |

*Cases:(1) $\mu_{annual} = 20\%$ and $\sigma_{annual} = 30\%$; (2) $\mu_{annual} = 40\%$ and $\sigma_{annual} = 45\%$,, and; (3) $\mu_{annual} = 80\%$ and $\sigma_{annual} = 60\%$
**(a) the high frequency data is uncorrelated; (b) the high frequency data is autocorrelated

In general, the VF-ARMA has optimal performance in a medium length series, i.e.: T = 510. In a longer series, it can potentially encounter convergence issues. It can also provide reasonable predictions when the high frequency data is exponential in form on the condition that weight of the covariates is low.

## 4.2. Varying Frequency in the Variance

Table 4 shows the average root-mean-square error and mean absolute deviation for all scenarios when $\mu_{annual} = 20\%$ and $\sigma_{annual} = 30\%$ and $\psi = 50\%$. The VF-GARCH yields the lowest average RMSE and MAD compared to the benchmark models across all scenarios. Unlike in VF-ARMA, failure of convergence is not an issue with VF-GARCH.

However, the VF-GARCH has several issues of its own. As shown in Table 6, the performance of the model decreases as the length of the series increases. As a consequence, the VF-GARCH is best suited for a short series, i.e.: T=255. The proposed model is also not suitable when the high frequency data is exponential in form.

**Table 4. RMSE and MAD for VF-GARCH and Benchmark Models**

| | | T | VF-GARCH | | GARCH | | GJR | |
|---|---|---|---|---|---|---|---|---|
| | | | RMSE | MAD | RMSE | MAD | RMSE | MAD |
| Linear Form | (a)* | 255 | 0.2999 | 0.0239 | 0.4015 | 0.0323 | 0.4015 | 0.0323 |
| | | 510 | 0.3632 | 0.0294 | 0.4194 | 0.0336 | 0.4194 | 0.0336 |
| | | 1530 | 0.4132 | 0.0330 | 0.4250 | 0.0339 | 0.4250 | 0.0339 |
| | (b)* | 255 | 0.4989 | 0.0399 | 0.6833 | 0.0550 | 0.6833 | 0.0550 |
| | | 510 | 0.5885 | 0.0470 | 0.7097 | 0.0568 | 0.7097 | 0.0568 |
| | | 1530 | 0.6763 | 0.0539 | 0.7160 | 0.0571 | 0.7160 | 0.0571 |
| Exponential Form | (a)* | 255 | 2.8192 | 0.2187 | 4.5495 | 0.3660 | 4.5495 | 0.3660 |
| | | 510 | 4.0587 | 0.3196 | 5.1511 | 0.4100 | 5.1511 | 0.4100 |
| | | 1530 | 4.6430 | 0.3680 | 5.7437 | 0.4583 | 5.7437 | 0.4583 |
| | (b)* | 255 | 2.9762 | 0.2289 | 5.3220 | 0.4272 | 5.3220 | 0.4272 |
| | | 510 | 4.1848 | 0.3323 | 5.5158 | 0.4428 | 5.5158 | 0.4428 |
| | | 1530 | 4.7550 | 0.3778 | 5.7625 | 0.4604 | 5.7625 | 0.4604 |

*For cases where $\mu_{annual} = 20\%$ and $\sigma_{annual} = 30\%$, $\psi = 50\%$*
*(a) the high frequency data is uncorrelated; (b) the high frequency data is autocorrelated

In general, the VF-GARCH has an optimal performance in a short series, i.e.: T = 255. Although it does not have any problems when it comes to convergence, it performs best in high frequency data that is linear in form.

Table 5 shows the results of validating the proposed models using out-of-sample forecasts. The series was forecasted 4 time points ahead, both VF-ARMA and VF-GARCH have better predictive ability than the benchmark.

**Table 5. MdAPE of the Proposed Models and Benchmark Model**

| T | Linear | | Exponential | |
|---|---|---|---|---|
| | Uncorrelated | Autocorrelated | Uncorrelated | Autocorrelated |

| | 255 | 74.98% | 72.20% | 65.03% | 67.50% |
| VF-ARMA | 510 | 84.35% | 83.22% | 84.90% | 82.45% |
| | 1530 | 90.02% | 87.42% | 91.75% | 91.18% |
| | 255 | 92.75% | 89.90% | 92.10% | 91.87% |
| VF-GARCH | 510 | 92.79% | 92.83% | 95.32% | 94.83% |
| | 1530 | 93.63% | 92.46% | 97.88% | 97.32% |
| | 255 | 107.50% | 105.41% | 101.26% | 104.15% |
| GARCH/GJR | 510 | 106.33% | 105.63% | 102.39% | 101.71% |
| | 1530 | 100.24% | 108.32% | 101.56% | 100.34% |

*For cases where $\mu_{annual} = 20\%$ and $\sigma_{annual} = 30\%$ and $\psi = 50\%$*

The predictive ability of VF-ARMA improves in a shorter series. Its predictive ability was shown to be optimal when the high frequency data is exponential. The predictive ability of the VF-GARCH is optimal when high frequency data is linear and autocorrelated. Simulations have shown that the proposed models can provide good estimates even with dependencies present in the high frequency data and, in the case of the VF-ARMA, even if the functional form of the high frequency data is exponential.

## 5. Application on Financial Data

The empirical feasibility of the proposed models and the estimation algorithms are used in modeling volatility to characterize stock returns in data reported at the weekly and daily levels. The data consists of a total of 520 observations at the daily level and 104 observations at the weekly level.

*Price*

The data used here is the daily stock data on Ayala Land, Inc. (ALI) for the period 2016-2017. All data was downloaded from the website of Finance Manila: (https://fm.advfn.com/). The data consists of the daily closing prices and the amount of volume traded per day. Figure 4 shows the price charts of ALI at the weekly and daily levels respectively.

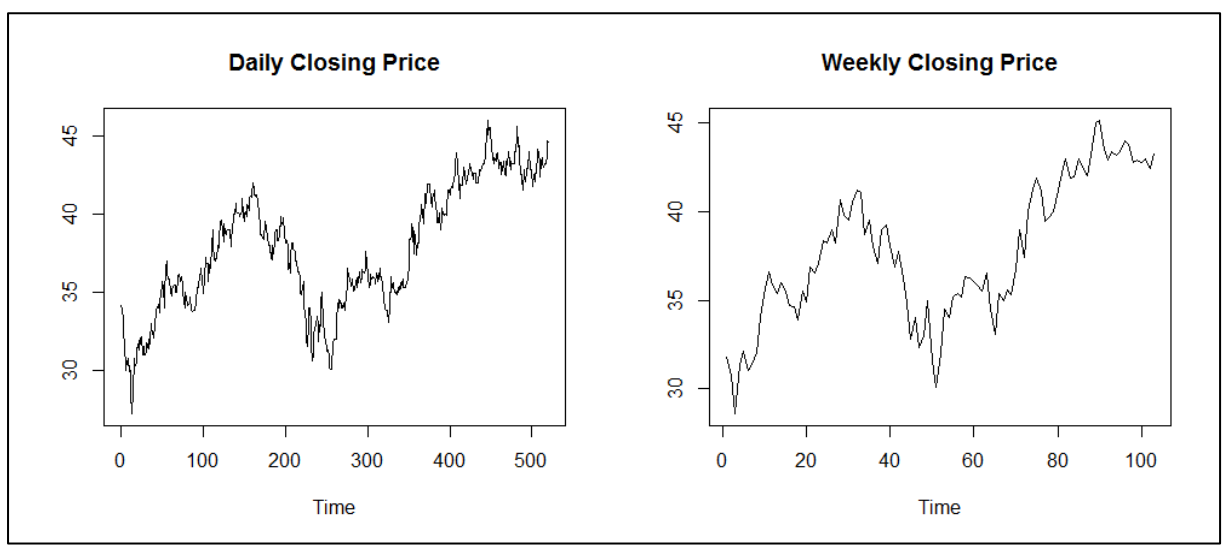

**Figure 4. Price Chart of ALI, January 2016 – December 2017**

Price at this period exhibited an upward trend, with the exception of the last quarter of 2016 (August 2016 – December 2016) where price was decreasing. Price at this period closed at Php 27.20/share and reached as high as Php 46.00/share.

*Trading Volume*

ALI, as a blue-chip company, i.e., one of the thirty companies indexed in the Philippine Stock Exchange Index, is one of the most frequently traded companies. Figure 5 shows that an average of 57 million shares were traded per week in this period, with the highest number of shares traded reaching as high as 243 million.

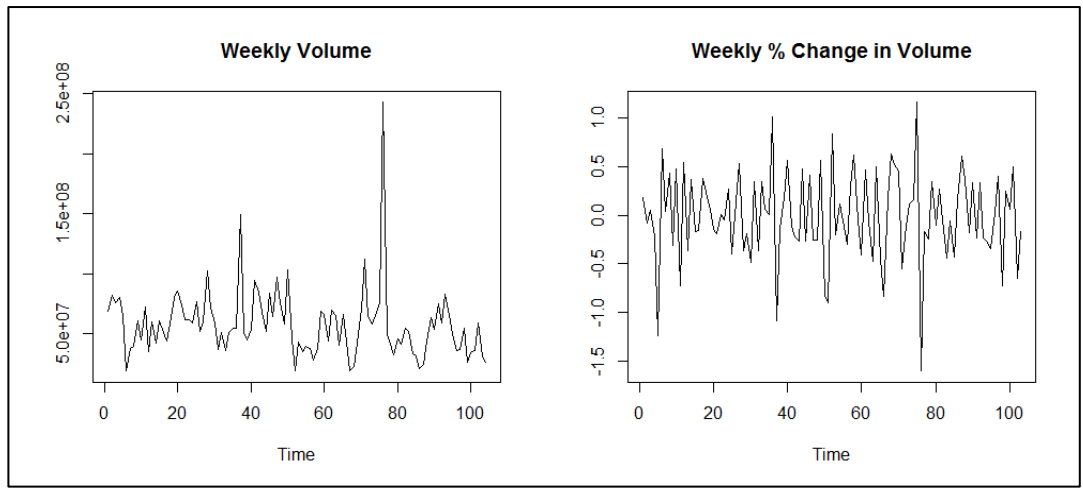

**Figure 5. Weekly Trading Volume of ALI and its Percentage Change, January 2016 – December 2017**

*Estimated Models*

The form of the semiparametric volatility models for stock returns estimated are given in Equations (15) and (16)

**VF-ARMA:**

$$\phi_p(B)y_t = \theta_q(B)a_t + \psi(B)v_t + \sum_{i=1}^{5} f_i(x_{i,t-1}) \qquad (15)$$

**VF-GARCH**

$$\sigma_t^2 = \omega + (\alpha + \gamma I_{t-1})a_{t-1}^2 + \beta\sigma_{t-1}^2 + \sum_{i=1}^{5} f_i(x_{i,t-1}) \qquad (16)$$

where  $y_t$ is the stock return at week t
  $v_i$ is the weekly percentage change in volume at week t
  $f(\cdot)$ is a continuous function in $x_{i,t-1}$
  $x_{i,t-1}$ is the stock price at day i of the previous week

For validation, the data was partitioned into a training and test data set. The training data set consisted of the first 99 observations while the test dataset consisted of the last four weekly observations, corresponding to approximately one trading month. The performance of the models for stock returns were assessed by computing the RMSE and MAD. The models were also compared to the standard GARCH and GJR-GARCH models just like in the simulations. The models were then validated using out-of-sample forecasts. The series was forecasted four steps ahead, the forecasted values were then compared to the test data set. The models were then evaluated using the MdAPE. The results are summarized in Table 6.

**Table 6. RMSE and MAD of Stock Returns Estimation**

|          | RMSE   | MAD    | MdAPE  |
|----------|--------|--------|--------|
| VF-ARMA  | 0.1036 | 0.0093 | 72.71% |
| VF-GARCH | 0.3433 | 0.0269 | 87.29% |
| GARCH    | 0.3488 | 0.0275 | 87.23% |
| GJR      | 0.3488 | 0.0275 | 87.23% |

VF-ARMA and VF-GARCH outperformed the benchmarks in in-sample predictive ability. Just like in the simulations, forecasts from VF-ARMA have the lowest reported errors, while errors from the VF-GARCH are comparable to those of the benchmarks.

For the out-sample predictive ability, the VF-ARMA reported a MdAPE of 72%, outperforming the benchmark at 87.23%. Meanwhile the predictive ability of the VF-GARCH was comparable to the benchmark models.

## 6. Conclusion

The semiparametric volatility models with varying frequencies are capable of generating better predictive ability compared to standard parametric GARCH and GJR models with aggregated high frequency variables. Aggregation indeed resulted to loss in information about the dynamic system that typically characterizes high frequency data.

The VF-ARMA in particular yield estimates with significantly lower errors as the length of a time series increases. Predictive ability was optimal even in short time series data and is superior to the parametric GARCH and GJR models. Furthermore, VF-ARMA is fairly robust as it was able to provide good estimates even when dependencies were present in the high frequency data and also when it was functionally exponential in form. Robustness was also highlighted in terms of its predictive ability as it is still able to provide good forecasts regardless of the presence of dependencies or functional form of the high frequency data.

The estimation procedure for VF-ARMA may not converge when the time series is very long. VF-GARCH on the other hand did not have problems in achieving convergence. Although it still yield better estimates than its parametric counterparts, its performance declines as the length of the series increases. Simulations have shown that its predictive ability was superior to the parametric GARCH and GJR models.

The proposed estimation in fitting a volatility model for stock returns are capable of producing good estimates for financial models given data of varying frequencies. With the accessibility of financial data online, the VF-ARMA can help traders make investment decisions that are data-driven and model-based as opposed to basing it purely on subjective analysis.